# Evaluating the Inclusiveness of Artificial Intelligence Software in Enhancing Project Management Efficiency – A Review

Vasileios Alevizos[1]
Karolinska Institute

Ilias Georgousis[1]
International Hellenic University

Akebu Simasiku
Zambia University College of Technology

Sotiria Karypidou
International Hellenic University

Antonis Messinis
Hellenic Electricity Distribution Network Operator (HEDNO)

November 18, 2023

## ABSTRACT

The escalating integration of Artificial Intelligence (AI) in various domains, especially project management, has brought to light the imperative need for inclusivity in AI systems. This paper investigates the role of AI software in augmenting both the inclusiveness and efficiency within the realm of project management. The research pivots around specific criteria that define and measure the inclusiveness of AI in project management, highlighting how AI, when developed with inclusiveness in mind, can significantly enhance project outcomes. However, there are inherent challenges in achieving this inclusivity, primarily due to biases embedded in AI learning databases and the design and development processes of AI systems. The study offers a comprehensive examination of AI's potential to revolutionize project management by enabling managers to concentrate more on people-centric aspects of their work. This is achieved through AI's ability to perform tasks such as data collection, reporting, and predictive analysis more consistently and efficiently than human counterparts. However, the incorporation of AI in project management extends beyond mere efficiency; it represents a paradigm shift in the epistemology of project management, calling for a deeper understanding of AI's role and impact on society. Despite these advantages, the adoption of AI comes with significant challenges, particularly in terms of bias and inclusiveness. Biased AI learning databases, which use shared and reusable datasets, often perpetuate initially discriminatory algorithms. Moreover, unconscious biases and stereotypes of AI designers, developers, and trainers can inadvertently influence the behavior of the AI systems they create. This necessitates a paradigmatic shift in how AI systems are developed and governed to ensure they do not replicate or exacerbate existing social inequalities. The research proposes a methodological approach involving the development of criteria for inclusion and exclusion, alongside data extraction, to evaluate the inclusiveness and efficiency of AI software in enhancing project management. This approach is crucial for understanding and addressing the challenges and limitations of AI in the context of project management. By focusing on inclusiveness, the study underscores the importance of a synergy between technological advancement and ethical consideration, demanding a comprehensive understanding and regulation to mitigate risks and maximize benefits. In conclusion, this paper presents a detailed exploration of AI's role in project management, highlighting both its potential benefits and the ethical challenges it poses. The findings and recommendations of this study contribute to the growing discourse on the need for inclusive AI systems in project management, offering insights for AI developers and project managers alike.

---

[1] Corresponding authors.

# 1. Introduction

In the past two decades, AI has increasingly become a fundamental aspect [Taboada, Daneshpajouh, Toledo, and De Vass(2023), Agrawal, Gans, and Goldfarb(n.d.)] of both personal and professional spheres. This trend is particularly evident in the domain of project management [Walker and Lloyd-Walker(2019)], where a paradigm shift towards automation for expedited feedback is becoming essential. The incorporation of AI in project management is not just about efficiency; it represents a significant change in the epistemology of project management. Evidently, AI's role in project management [Walker and Lloyd-Walker(2019)] is exemplified by its ability to perform tasks such as data collection, reporting, and predictive analysis more consistently and efficiently than human counterparts, as highlighted in recent literature. This emerging trend underscores the potential for AI to revolutionize project management by allowing managers to focus more on people-centric aspects of their work [Sravanthi, Sobti, Semwal, Shravan, Al-Hilali, and Bader Alazzam(2023)]. However, the adoption of AI comes with its own set of challenges [Shams, Zowghi, and Bano(2023), Cachat-Rosset and Klarsfeld(2023)], notably concerns regarding bias and inclusiveness. Recent initiatives [Cachat-Rosset and Klarsfeld(2023), Walsh, Levy, Bell, Elliott, Maclaurin, Mareels, and Wood(n.d.)] aimed at ensuring fair AI systems (AISs) have involved the creation of more diverse AI learning databases and the establishment of ethical principles and guidelines focusing on diversity, equity, and inclusion (DEI). These efforts are indicative of a growing awareness and urgency within the international community to regulate AI practices, especially to prevent negative impacts on vulnerable groups like women and racial minorities. Yet, the concrete composition and operational recommendations of these principles for DEI remain elusive. Fundamentally, the risk of discrimination and unfair treatment in AI predominantly [Shams, Zowghi, and Bano(2023), Cachat-Rosset and Klarsfeld(2023)] stems from three sources. Firstly, biased AI learning databases, which use shared and reusable datasets, often perpetuate initially discriminatory algorithms. Secondly, the unconscious biases and stereotypes of AI designers, developers, and trainers can inadvertently influence the behavior of the AISs they create. Thirdly, feedback from internet information or users [Ntoutsi, Fafalios, Gadiraju, Iosifidis, Nejdl, Vidal, Ruggieri, Turini, Papadopoulos, Krasanakis, Kompatsiaris, Kinder-Kurlanda, Wagner, Karimi, Fernandez, Alani, Berendt, Kruegel, Heinze, Broelemann, Kasneci, Tiropanis, and Staab(2020), P. S. (2023)], which may be biased and uncontrollable, significantly contributes to the propagation of bias in AI systems. This feedback, often idempotent in nature, comes from a myriad of online sources and user interactions that AI systems are exposed to. It presents an autonomic and ambient aspect of AI learning, making it a peripatetic challenge in ensuring AI fairness and inclusiveness. This user-generated content, rife with personal opinions, prejudices, and potentially misleading information, becomes an eigenvalue in the complex equation of AI ethics, further complicating the already intricate praxis of creating unbiased AI. These issues highlight the need for a paradigmatic shift in how AI systems are developed and governed, to ensure they do not replicate or exacerbate existing social inequalities. Together with these developments in AI and its implications for project management, a synergy between technological advancement and ethical consideration is crucial. This necessitates a pedagogical shift in how AI is integrated into professional practices, demanding a deeper exegesis of AI's role and impact on society. In conjunction with the advancement of AI technologies, there is a significant need [Wirtz, Weyerer, and Kehl(2022)] for comprehensive understanding and regulation to mitigate risks and maximize benefits. This juxtaposition of AI's potential in transforming project management and the ethical challenges it presents forms a complex, yet fundamentally important, area of contemporary discourse.

In this paper, an investigation is conducted to evaluate the inclusiveness and efficiency of AI software in enhancing project management. The objective is to assess the role of AI software in augmenting both inclusiveness and efficiency within the realm of project management. Central to this investigation are several research questions: the contribution of AI software to project management efficiency, the impact of AI on aspects of inclusiveness within project management, and the measurable outcomes of employing AI in project management scenarios. A methodological approach involving the development of criteria for inclusion and exclusion, alongside data extraction, is suggested. Challenges and limitations of AI in the context of project management are also discussed. AI pervades numerous facets of our daily lives, yet the opacity of current AI algorithms raises concerns regarding their trustworthiness and adoption. This opacity refers to the inability of AI systems to elucidate their inner workings or logic to users. Efforts to enhance the transparency of AI systems have led to the study of explainable AI (XAI), capturing the attention of the human-computer interaction community, who advocate for a human-centered approach to developing reliable, safe, and trustworthy AI applications. Notably, explainability in AI has been shown to influence user trust and subsequent adoption of AI systems. However, understanding how people interpret AI explanations and comprehend their causal relationships remains a less explored area. Trust in AI is a complex [Ntoutsi, Fafalios, Gadiraju, Iosifidis, Nejdl, Vidal, Ruggieri, Turini, Papadopoulos, Krasanakis, Kompatsiaris, Kinder-Kurlanda, Wagner, Karimi, Fernandez, Alani, Berendt, Kruegel, Heinze, Broelemann, Kasneci, Tiropanis, and Staab(2020), Shams, Zowghi, and Bano(2023)], multi-dimensional phenomenon influenced by contextual factors and extends beyond logical reasoning. This research turns to normative ethics, encompassing utilitarian, deontological, and virtue ethical perspectives, to broaden the understanding of trust in AI [Shams, Zowghi, and Bano(2023)]. These perspectives reveal different factors influencing trust in AI within specific contexts. The research aims to investigate how individuals comprehend AI decisions from a causality standpoint and interpret their interaction with AI in varied socio-technical contexts based on these ethical perspectives. The study aims to identify factors contributing to distrust or enhancing trustworthiness in AI solutions, providing insights for designers to increase AI systems' trustworthiness, along with discussions on human-AI trust, trustworthy AI, XAI, causality, and normative ethics, are presented, offering a comprehensive view of the multifaceted nature of trust in AI.

2. Background

Project managers shoulder a multifaceted array of responsibilities and goals, central to which is the effective coordination and completion of projects within the stipulated timeframes and budgets. Central to their role is the successful coordination and completion of projects, which involves intricate planning, team management, resource allocation, and problem-solving. Inasmuch as these tasks are challenging, the integration of inclusiveness within project management practices is increasingly being recognized as essential. Inclusiveness not only fosters a more harmonious and effective working environment but also ensures that diverse perspectives and skills are utilized, enhancing the overall quality and success of projects. The utilization of AI tools in project management has been profoundly affected by the need for inclusiveness. AI tools, if not designed with inclusivity in mind, risk reinforcing biases and excluding important stakeholder perspectives. Ergo, it is crucial that AI tools used in project management are developed with a focus on inclusiveness, ensuring that they cater to a diverse range of needs and preferences. This approach not only aligns with ethical principles but also enhances the effectiveness and efficiency of project management. Moreover, Large Language Models (LLMs) have emerged as a significant aid in facilitating project management tasks [Vidgof, Bachhofner, and Mendling(2023), Liu, Jiang, Zhang, Liu, Zhang, Biswas, and Stone(2023)] while also promoting inclusiveness. These models, through their advanced natural language processing capabilities, can automate several routine tasks, thereby freeing project managers to focus on more strategic aspects of their projects [Prieto, Mengiste, and García De Soto(2023), Pu, Yang, Li, Guo, and Li(2023)]. For instance, automated summarization of meetings by LLMs ensures that key points and decisions are concisely captured, making information accessible to all team members, including those who may have language or hearing barriers. This contributes to a more inclusive environment by ensuring that everyone is on the same page. Besides summarizing meetings, LLMs can also function as co-pilots in writing proposals. By generating drafts based on initial inputs, these models save time and reduce the workload on project managers, allowing them to dedicate more attention to strategic planning and team management. Not only do these AI tools streamline administrative tasks, but they also assist in ensuring that the language used is inclusive and free of biases, thereby promoting a culture of inclusiveness within project teams. In addition to these functions, LLMs can be programmed to suggest meeting hours that are convenient for all team members, considering different time zones and personal schedules. This feature is particularly valuable in today's globalized work environment, where teams are often spread across various geographical locations. By facilitating the scheduling of meetings that accommodate everyone, LLMs contribute to creating an inclusive and respectful working environment. Furthermore, the assignment of tasks by LLMs can be optimized to consider the unique strengths and preferences of team members, thereby ensuring that tasks are allocated fairly and effectively. This approach not only improves the efficiency of the project but also ensures that team members feel valued and included, fostering a positive team dynamic. Inasmuch as these AI tools offer numerous advantages, it is important to be mindful of potential limitations and challenges. Barring any unforeseen technological limitations or ethical concerns, the implementation of LLMs in project management could significantly transform how projects are planned, executed, and monitored. Therefore, it is incumbent upon project managers to stay abreast of these technological advancements and integrate them thoughtfully into their project management practices. The integration of AI tools, particularly LLMs, into project management not only streamlines various tasks but also plays a crucial role in promoting inclusiveness. By automating routine tasks such as meeting summarization, proposal drafting, meeting scheduling, and task assignment, LLMs enable project managers to focus on more strategic aspects of project management while fostering an inclusive and efficient working environment.

2.1. Inclusive AI

Inclusive AI, a concept of paramount importance in contemporary technological development, refers to the development of AI systems that are equitable and fair to all users [Ntoutsi, Fafalios, Gadiraju, Iosifidis, Nejdl, Vidal, Ruggieri, Turini, Papadopoulos, Krasanakis, Kompatsiaris, Kinder-Kurlanda, Wagner, Karimi, Fernandez, Alani, Berendt, Kruegel, Heinze, Broelemann, Kasneci, Tiropanis, and Staab(2020)], irrespective of their demographic characteristics. This concept is crucial for ensuring that AI systems do not perpetuate or amplify existing societal biases. In the context of project management, the importance of inclusive AI is underscored by the need to avoid biases that could lead to suboptimal decision-making and unfair practices. The need for inclusivity in AI arises from a recognition of the diverse range of biases that can influence project planning and management. Strategic misrepresentation, a form of bias, is characterized by the deliberate distortion of information for strategic benefits. This is akin to political or power bias, where information is manipulated to gain advantage. Optimism bias, another prevalent tendency, involves an overestimation of positive outcomes and underestimation of negatives, leading to unrealistic project expectations. Similarly, the uniqueness bias leads project managers to view their projects as more exceptional than they are, affecting judgment and decision-making. The planning fallacy, manifested as underestimating costs and risks while overestimating benefits, along with overconfidence bias, where excessive confidence is placed in one's own judgments, can significantly skew project outcomes. Hindsight bias, or the I-knew-it-all-along effect, affects how past events are perceived, often as more predictable than they were. Availability bias leads to overestimating the likelihood of easily recalled events, whereas base rate fallacy involves ignoring general information in favor of specifics. Anchoring, or the reliance on initial information, and escalation of commitment, where decision-makers justify increased investment despite contrary evidence (sunk cost fallacy), further exemplify biases in project management. Real-world examples of bias in AI systems are numerous and varied. In the criminal justice system, the COMPAS system exhibited biases against African-American defendants [Ferrara(2023)], marking them as higher risk without prior convictions. In healthcare, an AI system predicting patient mortality rates showed biases against African American patients, potentially leading to inadequate treatment. Similarly, facial recognition technology, found to be less accurate for people with darker skin tones, raises concerns about wrongful legal implications. Types of biases in AI systems include sampling bias, where training data is unrepresentative of the target population, leading to biased predictions. Algorithmic bias results from design choices that prioritize certain attributes, often leading to unfair outcomes. Representation bias occurs when datasets do not accurately model the population, resulting in inaccurate predictions. Confirmation bias arises when AI systems reinforce pre-existing beliefs of their creators or users. Measurement bias, involving systematic over- or underrepresentation in data collection, and interaction bias, where AI systems treat users unfairly based on human interaction, are additional concerns. The importance of inclusive AI in project management lies in its capacity to mitigate these biases. Through ensuring representative and equitable AI systems, project managers can make more accurate, fair, and effective decisions. Inclusive AI promotes not only ethical responsibility but also enhances the effectiveness and reliability of AI systems in diverse environments. Furthermore, inclusive AI aligns with normative ethical principles, contributing to a more equitable technological landscape. Additionally, the development of inclusive AI involves embracing positivism and constructivism, acknowledging the mutable nature of AI and its capacity to be shaped by human values and perspectives. The ontology of AI systems, defined by their inherent characteristics and behaviors, must be scrutinized to ensure they align with inclusive principles. Similarly, the application of syllogism in AI development ensures logical consistency in decision-making processes, reducing the likelihood of bias. Nonetheless, the pursuit of inclusive AI in project management is not without challenges. The idiosyncrasies of various AI systems, coupled with the ambient complexities of real-world applications, necessitate a thorough understanding of the potential biases and their impacts. The immutable nature of certain algorithmic processes, once established, poses a challenge to the dynamic nature of inclusivity efforts. Ensuring that AI systems are both inclusive and efficient requires a careful balancing act, considering the diverse ensemble of stakeholders involved.

### 2.1.1. Trust of unexplainable AI

XAI, an emerging concept within the realm of artificial intelligence, signifies the development of AI systems [Esmaeili, Vettukattil, Banitalebi, Krogh, and Geitung(2021)] that are transparent and comprehensible in their operations. It encompasses methodologies and technologies that elucidate the functioning of AI models, making their processes and outcomes clear and understandable to users. As described by Rudresh Dwivedi and Devamanyu Hazarika [Dwivedi, Dave, Naik, Singhal, Omer, Patel, Qian, Wen, Shah, Morgan, and Ranjan(2023)], XAI is a burgeoning field that aims to enhance the comprehensibility of ML models for human users, thereby facilitating a better understanding, appropriate trust, and the development of more interpretable models. This aspect of AI is particularly crucial in project management, where decision-making processes and outcomes must be transparent and justifiable. In fact, XAI can be perceived as a critical attribute of AI systems, aiming to reduce their inherent opacity and thereby enhance their trustworthiness. The concept of XAI aligns with definitions proposed in the literature, where it is described [Esmaeili, Vettukattil, Banitalebi, Krogh, and Geitung(2021), Wenninger, Kaymakci, and Wiethe(2022)] as the capability of an AI system to provide details or reasons for its functioning in a manner that is easily comprehensible. This understanding of explainability and XAI is pivotal for making AI systems less opaque and more user-friendly. However, the exploration of how individuals interpret AI explanations and comprehend their causal relationships remains a significant research area [Ntoutsi, Fafalios, Gadiraju, Iosifidis, Nejdl, Vidal, Ruggieri, Turini, Papadopoulos, Krasanakis, Kompatsiaris, Kinder-Kurlanda, Wagner, Karimi, Fernandez, Alani, Berendt, Kruegel, Heinze, Broelemann, Kasneci, Tiropanis, and Staab(2020), Ferrara(2023)]. This understanding is referred to as 'castability,' which denotes the degree to which an explanation facilitates a user's causal understanding effectively, efficiently, and satisfactorily in a specific context. Castability, being critical for achieving human-understandable explanations, has been shown to positively impact explainability and, in tandem, build user trust towards AI systems. It plays a crucial role in increasing the trustworthiness of AI by revealing key signals in the AI reasoning process [Esmaeili, Vettukattil, Banitalebi, Krogh, and Geitung(2021)], allowing users to anticipate behavior in risky situations, and conversely, increasing distrust in non-trustworthy AI by highlighting when desired behaviors are not manifested. For project managers, the incorporation of XAI techniques in AI solutions is indispensable for fostering the adoption of these systems by individuals and organizations. The relationship between XAI and user trust has been a focus of recent studies, underscoring the necessity of XAI in building trust. Nevertheless, there is a scarcity of research specifically investigating the effect of castability on explainability, indicating a need for further exploration in this area. Trust, a crucial element for AI systems, is often predicated on the belief that people and organizations are more likely to cooperate with, use, and adopt systems they trust. Philosophically, trust in AI systems is grounded in normative ethics, which investigates how AI systems ought to act to convince people of their trustworthiness. Normative ethics encompasses three classical schools of thought—utilitarian, deontological, and virtue ethical—each providing different perspectives and conflicting arguments on how people understand AI, its decisions, and the surrounding socio-technical context. In the realm of project management, XAI is essential for ensuring transparency, interoperability, and reliability of AI systems. Heuristic approaches, which are often subject to cognitive biases, can benefit significantly from XAI techniques. These techniques support rational reasoning processes and target decision-making errors, thereby aligning with the principles of utilitarianism in promoting positive outcomes. Specifically, in project management, XAI enables managers to make informed decisions based on clear, understandable AI reasoning, thereby enhancing the efficacy and fairness of project outcomes.

2.1.2. AI Decision Transparency

The burgeoning field of AI has ushered in a new era of decision-making tools, with a specific emphasis on AI Decision Transparency [Kim, Park, and Suh(2020), De Fine Licht and De Fine Licht(2020)]. This concept is critical in various sectors, notably in project management. AI Decision Transparency involves making the decision-making processes of AI systems clear, understandable, and accountable to users and stakeholders. It addresses the growing concern that AI systems, especially LLMs like AutoRepo [Pu, Yang, Li, Guo, and Li(2023)], can make decisions or generate outputs that are not easily interpretable or may even be erroneous. The example of AutoRepo, an automated framework for generating construction inspection reports, encapsulates the potential and challenges of AI in decision-making processes. The framework utilizes unmanned vehicles and LLMs to collect and process scene information, thereby automating the creation of compliant and standardized construction inspection reports. However, hereinbefore lies the critical aspect of responsibility in AI deployment and usage. Insofar as AI systems like AutoRepo streamline and enhance efficiency in complex tasks, they also bring to the fore the issue of potential mistakes and 'hallucinations'[Dziri, Milton, Yu, Zaiane, and Reddy(2022), Ji, Lee, Frieske, Yu, Su, Xu, Ishii, Bang, Madotto, and Fung(2023)]—instances where AI systems generate inaccurate or misleading information. Notwithstanding the benefits of such systems, the responsibility for these AI-induced errors must be de facto placed upon the authors and developers of these AI systems. Albeit AI tools can significantly aid in research and project management, their fallibility necessitates a vigilant and accountable approach from those who implement them. Hence, authors and developers must bear the primary responsibility for ensuring the integrity and accuracy of the outputs generated by AI systems. This obligation is paramount [Giray(2023), Shams, Zowghi, and Bano(2023)], especially considering the iterative nature of AI learning and the convolutional complexities involved in machine learning models. It requires a comprehensive understanding of the dataset used, the optimization techniques applied, and the potential biases embedded within the AI system. The ensemble of these factors contributes to the overall reliability and trustworthiness [Athaluri, Manthena, Kesapragada, Yarlagadda, Dave, and Duddumpudi(2023)] of AI-generated decisions and outputs [Prieto, Mengiste, and García De Soto(2023)]. In the context of project management, where AI tools like AutoRepo are increasingly utilized for tasks such as automated summarization of meetings, proposal writing, and task allocation, the importance of decision transparency cannot be overstated. AI tools should be designed and used with a clear understanding of their limitations and potential errors. This approach ensures that project managers can rely on AI tools for efficiency without compromising the quality and accuracy of the information provided, thereby maintaining the trust and confidence of all stakeholders involved.

## 2.2. Adaptation of Unbiased Autonomous Project Manager

The adaptation of unbiased autonomous project managers, particularly with AI tools like ChatGPT, represents a significant development in the field of project management [VUSUMUZI MAPHOSA and MFOWABO MAPHOSA(2022)]. This approach is beneficial for several reasons. Firstly, AI tools can automate routine tasks, thus saving time and reducing human error [Athaluri, Manthena, Kesapragada, Yarlagadda, Dave, and Duddumpudi(2023)]. For instance, ChatGPT can efficiently handle scheduling, resource allocation, and progress tracking, thereby streamlining project management processes [Abbas, Younus, Malik, and Hassan(2023)]. This automation represents the trend of integrating AI into management practices to enhance efficiency. Secondly, these AI systems can provide data-driven insights that aid in decision-making [Kim, Park, and Suh(2020)]. By analyzing large datasets, AI tools like ChatGPT can uncover patterns and insights that might not be immediately apparent to human project managers, thereby enabling more informed decisions. However, the use of AI tools in project management also introduces certain challenges and dilemmas, particularly regarding biases inherent in AI models and datasets. These biases can potentially lead to skewed decision-making, contradicting the objective of unbiased project management. For example, if an AI tool has been trained on data that is not representative of the diverse demographics of a project's stakeholders, it may make decisions that favor certain groups over others. Additionally, these tools might not adequately account for the unique nuances of specific projects, leading to generic or inappropriate recommendations. Measuring AI inclusiveness becomes critical in this context. While this measurement can offer valuable insights into the performance [Ferrara(2023)] of AI tools across diverse groups and scenarios, it also presents its own set of challenges. The diversity of datasets and the complexity of AI algorithms can make accurately assessing and mitigating biases a difficult task. For instance, an AI tool might perform well in certain standard scenarios but fail to deliver the same level of performance in more complex or unusual situations. This variation in performance can be particularly challenging to measure and rectify. In the second part of this discussion, the focus shifts to the practical aspects of implementing unbiased AI in project management. The key challenge lies in ensuring that AI project management models avoid biases to ensure fair and effective decision-making. This requires regular updates and refinements of AI systems, incorporating feedback from diverse user groups. For example, if an AI model used in project management shows a tendency to allocate resources in a way that consistently overlooks certain teams or departments, it needs to be recalibrated to correct this bias. Improvements in AI project management models should focus on enhancing the fairness and representativeness of the datasets used. Iterative updates and refinements, alongside incorporating diverse perspectives in model development and training, can aid in mitigating biases. For instance, including data from a wide range of project types and demographics can help ensure that the AI tool's recommendations are applicable and fair across various scenarios. Moreover, regular audits and assessments of AI decisions can help identify and address any emerging biases. Furthermore, there is a need for collaboration between AI developers and project management professionals to ensure that the AI tools are tailored to the specific needs and challenges of project management. This collaboration can take the form of joint workshops, feedback sessions, and pilot projects to test and refine AI tools in real-world project management scenarios. By working closely with those who will use these tools daily, developers can gain valuable insights into the practical requirements and potential pitfalls of AI in project management. In the third part of this discussion, the future of unbiased autonomous project managers using AI is considered. The evidence suggests that there is significant potential for AI to revolutionize project management. However, realizing this potential requires addressing the current limitations and challenges. Furthermore, could lead to skewed decision-making, adversely affecting the inclusiveness and fairness of project outcomes. For example, AI-driven resource allocation decisions might inadvertently favor certain groups or project aspects, resulting in imbalanced development. Additionally, there exists the issue of over-reliance on AI recommendations, which might discourage critical thinking and independent decision-making among project managers. In essence, while AI tools can offer valuable suggestions, they should not replace human judgment and experience. Future developments in this area could involve more advanced AI algorithms capable of learning and adapting over time. For example, AI systems might use machine learning to continuously improve their decision-making processes based on feedback and outcomes from past projects. This self-improving capability could help AI tools become more accurate and fairer in their recommendations and decisions. Another area of future development is the integration of AI tools with other technologies. For example, combining AI with virtual reality (VR) or augmented reality (AR) could create immersive project management environments where managers can visualize and interact with project data in new ways. This integration could make project management more intuitive and effective, allowing managers to identify potential issues and opportunities more quickly. While the adaptation of unbiased autonomous project managers using AI tools like ChatGPT presents promising opportunities for enhancing efficiency and decision-making in project management, it also brings challenges that must be carefully navigated. Addressing biases in AI models, ensuring the representativeness of datasets, and maintaining a close collaboration between AI developers and project management professionals are essential steps towards realizing the full potential of AI in project management. With continued development and refinement, AI has the potential to become an indispensable tool in the arsenal of modern project managers.

3. Measuring Inclusiveness in AI-Enhanced Project Management

In developing a methodology for measuring inclusiveness in AI-enhanced project management, it is essential to establish a comprehensive set of criteria that encompass various dimensions of AI integration and functionality. These criteria are pivotal for assessing how AI tools and systems impact diverse groups within the context of project management. The first dimension, technical aspects, includes criteria such as AI Decision Transparency, which emphasizes the need for clarity in how AI decisions are made. This is crucial in ensuring that stakeholders understand the rationale behind AI-driven decisions, fostering trust in the system. Algorithmic Bias Detection is another critical criterion, focusing on measures to identify and correct biases in algorithms. This is essential for preventing discriminatory outcomes and ensuring fairness in decision-making processes. Data Diversity plays a significant role in training AI systems, where the use of diverse datasets can prevent biases and improve the system's overall inclusivity. Additionally, Accessibility is a crucial criterion, ensuring that AI tools are accessible to users with disabilities, thereby promoting an inclusive work environment. Language Support, particularly the multilingual capabilities of AI tools, is essential in today's globalized project environments, enabling effective communication across diverse linguistic backgrounds. User interaction forms the second key area of focus, encompassing criteria like User Interface Usability, which evaluates the ease of use of the AI system. An intuitive and user-friendly interface is crucial for ensuring that all team members can effectively utilize the AI tool, irrespective of their technical expertise. Customization Options are also vital, allowing users to tailor AI tools to their individual needs and preferences, thereby enhancing the tool's applicability and effectiveness. Feedback Mechanisms for user input on AI performance are critical for iterative improvement of the system, ensuring that it evolves in response to user needs and experiences. Cultural Sensitivity is another important criterion, reflecting AI's awareness of and sensitivity to cultural differences. This is crucial for ensuring that AI tools are appropriately attuned to the varied cultural backgrounds of users. Additionally, User Education and Support are necessary to ensure users are well-equipped to utilize AI tools effectively, thereby maximizing the benefits of AI integration in project management. In the realm of project management, specific criteria are critical for evaluating the effectiveness of AI tools. Task Allocation Efficiency assesses AI's effectiveness in distributing tasks among team members, a crucial factor for project success. Resource Optimization evaluates AI's role in optimizing the use of project resources, crucial for maintaining project budgets and timelines. Risk Assessment Accuracy is another key criterion, determining AI's ability to accurately assess project risks and inform decision-making. Collaboration Enhancement evaluates how AI fosters team collaboration, an essential aspect of successful project management. Finally, Timeline Management assesses AI's effectiveness in managing project timelines, ensuring that projects stay on track and meet their deadlines. The second part of the methodology focuses on ethical considerations, performance metrics, and the organizational impact of AI-enhanced project management. Ethical considerations include criteria such as Privacy Protection, which assesses AI's adherence to user data privacy standards. This is crucial in maintaining the confidentiality and integrity of project-related data. Security Measures evaluate AI's features to protect user data from unauthorized access or breaches, an essential aspect in the digital age. Consent Protocols involve processes for obtaining user consent, particularly when dealing with sensitive data, ensuring compliance with legal and ethical standards. Transparency in Data Usage assesses the clarity on how user data is utilized by AI systems, fostering transparency and trust. Lastly, Ethical Standards Compliance evaluates AI's adherence to ethical guidelines and principles, ensuring that AI tools are used responsibly and ethically. Performance metrics are crucial for assessing the effectiveness of AI tools in project management. Accuracy in Predictive Analytics evaluates the precision of AI in forecasting project-related outcomes, a key factor in strategic planning. The Error Rate criterion assesses the frequency of errors in AI outputs, which is critical for reliability. Response Time evaluates the speed at which AI systems respond to user queries, impacting user satisfaction and efficiency. Scalability assesses AI's ability to handle varying project sizes and complexities, a vital feature for diverse project environments. Integration with Existing Systems evaluates the compatibility of AI tools with current project management tools, ensuring seamless integration and user experience. The organizational impact of AI in project management is measured through criteria such as Employee Satisfaction, which assesses the impact of AI on employee morale and engagement. Change Management evaluates the ease of integrating AI into existing workflows and processes, a critical aspect for successful adoption. Training Requirements assess the level of training needed for effective AI adoption, impacting the ease of integration and user competency. Cost-Benefit Analysis assesses the economic impact versus the benefits of AI implementation, crucial for justifying AI investments. Long-term Sustainability evaluates AI's viability for long-term usage, ensuring that the investment in AI tools yields enduring benefits. The third part of the methodology encompasses social and environmental considerations, innovation and development, and stakeholder involvement. Social and environmental considerations include criteria such as Social Impact, which assesses AI's broader impact on society, ensuring that AI tools contribute positively beyond the confines of project management. Environmental Sustainability evaluates AI's environmental footprint, promoting environmentally responsible AI solutions. Community Engagement assesses the involvement of AI tools with local communities, fostering social responsibility. Social Responsibility evaluates AI's contribution to social causes, aligning AI implementation with corporate social responsibility goals. Legal Compliance assesses adherence to relevant laws and regulations, ensuring that AI tools operate within legal frameworks. Innovation and development focus on criteria such as Research and Development Investment, which evaluates the resources allocated to improving AI tools, fostering continuous innovation. Innovative Features assesses the introduction of novel features in AI tools, enhancing their capabilities and user experience. Adaptability to Trends evaluates AI's responsiveness to market trends and evolving project management needs. Futureproofing assesses AI's capacity to evolve with technological advancements, ensuring long-term relevance. Cross-Industry Applications evaluate the versatility of AI tools across different industries, enhancing their applicability and value. Stakeholder involvement is crucial for the successful implementation and acceptance of AI in project management. Criteria such as Client Satisfaction assess how clients perceive the AI's impact on project outcomes. Stakeholder Communication evaluates AI's role in facilitating effective communication among project stakeholders. Vendor Relationships assess the impact of AI on relationships with suppliers and partners. Investor Interest evaluates the attraction of investors due to AI implementation, indicating the perceived value of AI in project management. Community Feedback assesses the feedback from the wider community on AI use, ensuring that AI tools are responsive to external perspectives and needs. These criteria are grouped into categories such as Technical and Functional Aspects, User Experience and Interaction, Ethical and Social Responsibility, Organizational and Economic Impact, and Stakeholder Engagement and Impact. This structured approach ensures a holistic assessment of AI inclusiveness in project management, considering all relevant factors from technical functionality to societal impact.

In the development of unbiased AI systems for project management, a comprehensive and nuanced approach is required, focusing on training general datasets and interactive systems to ensure inclusiveness and avoid biases [Ntoutsi, Fafalios, Gadiraju, Iosifidis, Nejdl, Vidal, Ruggieri, Turini, Papadopoulos, Krasanakis, Kompatsiaris, Kinder-Kurlanda, Wagner, Karimi, Fernandez, Alani, Berendt, Kruegel, Heinze, Broelemann, Kasneci, Tiropanis, and Staab(2020)]. This approach involves a meticulous selection and adjustment of various hyperparameters that influence the AI model's learning process and decision-making capabilities. Key hyperparameters include the learning rate, which determines the speed at which the AI model learns, and batch size, affecting how much data the model processes at one time. The number of epochs is pivotal in defining how many times the entire dataset is passed through the neural network. The depth and width of neural networks, determined by the number of layers and neurons, respectively, play a crucial role in the model's capacity to process complex data structures. Activation functions are essential in adding non-linearity to the model, allowing it to learn and perform more complex tasks. The choice of optimizer type and loss function directs the model's approach to reducing errors during training. Additionally, the dropout rate and regularization methods, such as L1 and L2, are critical in preventing overfitting, ensuring that the model generalizes well to new, unseen data. Initial weight settings and data normalization or standardization techniques lay the foundational structure for unbiased learning. Further, advanced parameters such as learning rate decay and momentum are employed to refine the model's learning process. For tree-based methods, parameters like the number of trees, max depth of trees, and min samples split play a significant role in the model's structure and decision process. Ensemble methods, such as bootstrapping and subsampling, enhance the model's robustness and accuracy by combining multiple models or subsets of data. In training interactive systems with live information, a diverse range of variables is incorporated to ensure the AI system's responsiveness and inclusiveness to various user profiles and contexts. These variables encompass user demographics, location data, behavior patterns, and real-time feedback, which enable the AI system to understand and cater to a wide array of user preferences and needs. The language and dialect used, along with device specifications, provide insights into user interaction modalities and preferences. Cultural context and ethical considerations are integrated to ensure that the AI system is sensitive to diverse cultural norms and adheres to ethical standards in user interactions. The second part of the methodology emphasizes the importance of inclusiveness and bias avoidance in AI systems used for project management. To achieve this, the AI model must be trained on diverse datasets, ensuring that it is representative of different user groups and scenarios. This involves considering variables like user demographics, behavior patterns, and real-time feedback. Such inclusiveness is vital to avoid biases that might arise from training on homogeneous datasets, which can lead to skewed decision-making and marginalization of certain user groups. In terms of user interaction, criteria like cultural sensitivity and language support are critical. AI systems must be designed to recognize and respect cultural differences and linguistic diversity, ensuring that they are accessible and relevant to users from various cultural backgrounds. Accessibility is another crucial aspect, where AI systems should be usable by people with disabilities, incorporating features like screen readers and voice commands. This ensures that the AI tools are inclusive and cater to a broad spectrum of users, enhancing user experience and satisfaction. Moreover, the methodology also involves the ethical considerations of AI system development. This includes ensuring privacy protection and security measures to safeguard user data. AI systems must adhere to consent protocols, obtaining explicit consent from users for data collection and usage. Transparency in how user data is utilized is essential for building trust and ensuring ethical AI practices. Ethical standards compliance ensures that AI tools are used responsibly, respecting user rights and societal norms. Performance metrics are crucial in evaluating the effectiveness and inclusiveness of AI tools. Accuracy in predictive analytics assesses the precision of AI in forecasting project-related outcomes. Error rate and response time are indicators of the AI system's reliability and efficiency. Scalability measures the AI tool's ability to handle varying project sizes and complexities, while integration with existing systems evaluates the tool's compatibility with current project management frameworks. The organizational impact of AI in project management is another essential aspect of the methodology. Employee satisfaction assesses the impact of AI on the workforce's morale and engagement. Change management evaluates the ease with which AI can be integrated into existing workflows and processes. Training requirements determine the level of instruction necessary for effective AI adoption. A cost-benefit analysis is crucial for justifying the investment in AI tools, assessing the economic impact versus the benefits. Long-term sustainability evaluates the AI tool's viability for continued use, ensuring that the investment in AI yields enduring benefits. In the third part of the methodology, social and environmental considerations, innovation and development, and stakeholder involvement are addressed. Social impact assesses the broader societal implications of AI tools in project management. Environmental sustainability evaluates the ecological footprint of AI systems, promoting environmentally responsible solutions. Community engagement measures the involvement of AI tools with local communities, fostering social responsibility. Legal compliance ensures that AI tools operate within legal frameworks, adhering to relevant laws and regulations. Innovation and development focus on the continuous improvement and evolution of AI tools. Research and development investment evaluates the resources allocated to enhancing AI systems. Innovative features assess the introduction of novel functionalities in AI tools. Adaptability to trends evaluates the AI tool's responsiveness to evolving market trends and project management needs. Futureproofing assesses the AI tool's capacity to evolve with technological advancements, ensuring its long-term applicability and relevance. Stakeholder involvement is crucial for the successful implementation and acceptance of AI in project management. Client satisfaction measures how clients perceive the AI's impact on project outcomes. Stakeholder communication assesses the AI tool's role in facilitating effective communication among project stakeholders. Vendor relationships evaluate the impact of AI on relationships with suppliers and partners. Investor interest gauges the attraction of investors due to AI implementation, indicating the perceived value of AI in project management. Community feedback assesses the broader community's response to AI use, ensuring that AI tools are responsive to external perspectives and needs.

3.1. Non-inclusive AI operations

While AI has the potential to significantly streamline and enhance project management processes, its effectiveness and ethicality hinge on its inclusivity. The blithe implementation of AI without consideration of inclusiveness leads to significant challenges, where AI operations may inadvertently propagate biases or exclude crucial stakeholder perspectives. Assiduous attention is thus required in monitoring and managing AI operations to ensure inclusiveness and avoid biases that could undermine the effectiveness of project management. Evaluating non-inclusive AI operations involves a structured analysis approach, where additional measures are considered and correlated through well-defined formulas. User satisfaction surveys, inclusiveness audits, and tracking usage patterns by diverse groups form part of the additional measures for assessing inclusiveness in project management AI. These measures, along with employee feedback mechanisms and the diversity of the AI training team, offer insights into how inclusively AI tools are being used and perceived. For training interactive systems with live data, measures such as data source diversity, algorithmic transparency reports, user control over data, ethical usage metrics, real-time bias monitoring, and inclusivity KPIs are indispensable. Each of these measures contributes to understanding and addressing potential biases in AI systems, ensuring that they operate in a manner that is inclusive and fair to all users. To effectively correlate these measures and derive meaningful insights, various types of analyses are employed. Descriptive analysis provides an overview of the dataset's basic features, while comparative analysis allows for comparisons of inclusiveness metrics across different time periods or user groups. Correlation analysis helps understand the relationships between different variables, such as user satisfaction and AI inclusiveness. Regression analysis predicts the impact of certain variables on inclusiveness outcomes, and factor analysis identifies underlying factors contributing to inclusiveness. Cluster analysis is used for segmenting users or projects based on inclusiveness metrics. The creation of a unified formula for inclusiveness involves assigning weights to each measure based on its importance and scoring each measure, which could be based on quantitative data or qualitative assessments. The Unified Inclusiveness Score (UIS) is then calculated by multiplying each score by its respective weight and summing them up. This formula, cognizant of the peculiarities of AI operations in project management, offers a laconic yet comprehensive approach to measuring inclusiveness.

*Table 1 Measures for Inclusiveness where Wn denote the weight assigned to each measure.*

| Measure | Description | Weight |
| --- | --- | --- |
| User Satisfaction Surveys | Regular surveys to gauge user contentment with AI tools. | W1 |
| Inclusiveness Audits | Reviews by external experts to assess the inclusiveness of AI applications. | W2 |
| Usage Patterns by Diverse Groups | Analysis of how different groups use the AI tool. | W3 |
| Employee Feedback Mechanism | Channels for employees to provide feedback on AI inclusiveness. | W4 |
| Diversity of AI Training Team | Ensuring that the team creating and managing AI systems reflects diverse perspectives. | W5 |
| Impact Assessments | Regular assessments of AI's impact on various groups. | W6 |
| Cultural Competence Training | Training for teams in cultural awareness and sensitivity for AI system development. | W7 |

The first table encapsulates key measures to ensure the inclusiveness of AI in project management, each weighted for its relative importance. User Satisfaction Surveys, represented by W1, involve indefatigable efforts to understand user contentment with AI tools through regular surveys. This provides perspicacious insight into how users perceive the AI's functionality and user-friendliness. Inclusiveness Audits, denoted by W2, juxtapose the AI's actual performance against inclusivity standards through reviews by external experts. These audits are crucial in identifying and addressing potential biases in AI applications. The measure of Usage Patterns by Diverse Groups, weighted as W3, examines how different user groups interact with the AI tool. This analysis can reveal esoteric usage patterns that might exacerbate biases if not addressed. For example, a particular AI feature might be predominantly used by one demographic group, leading to a skewed development focus. The Employee Feedback Mechanism, indicated by W4, opens channels for employees to voice their opinions on AI inclusiveness. This feedback is integral to continually refining the AI tool to meet diverse needs. Furthermore, the Diversity of the AI Training Team (W5) ensures that the team reflects a broad spectrum of perspectives. This diversity is a laudable goal as it prevents the ingraining of a homogeneous viewpoint in AI development. Impact Assessments, marked by W6, involve regular evaluations of AI's impact on various groups, ensuring that the AI does not inadvertently disadvantage any user group. Lastly, Cultural Competence Training (W7) involves training AI teams in cultural awareness and sensitivity, essential for developing AI systems that respect and adapt to cultural diversity. This measure is particularly important in global projects where cultural nuances significantly influence user interaction with technology.

*Table 2 Variables for Training Interactive Systems with Live Data, where Vn denote the weight assigned to each variable.*

| Variable | Impact on Inclusiveness | Weight |
| --- | --- | --- |
| Data Source Diversity | Ensures diverse data sources for AI training to avoid biases. | V1 |
| Algorithmic Transparency Reports | Provides transparency in AI decision-making processes. | V2 |
| User Control Over Data | Allows users to control their data usage in AI systems. | V3 |

| Variable | Impact on Inclusiveness | Weight |
|---|---|---|
| Ethical Usage Metrics | Metrics to evaluate the ethical use of live data in AI systems. | V4 |
| Real-time Bias Monitoring | Continuously monitors for biases in AI operations. | V5 |
| Inclusivity KPIs | Key Performance Indicators specifically related to inclusiveness. | V6 |

The table 2 delineates various measures for inclusiveness in Project Management AI, emphasizing the cumulative impact of diverse factors. Under User Experience, User Satisfaction Surveys and Employee Feedback Mechanism, for instance, are pivotal for gauging contentment with AI tools and gaining insights from those interacting directly with these systems. In the Diversity & Inclusion category, tracking Usage Patterns by Diverse Groups and ensuring the Diversity of the AI Training Team are crucial for maintaining a heterogeneous user and developer base, preventing the creation of AI systems tailored only for a homogeneous group. The Cultural Competence section, through measures like Cultural Competence Training, enhances AI's effectiveness across diverse cultural landscapes, addressing the needs that might otherwise be like finding a needle in a haystack for generic AI systems. Inclusiveness Audits in the Inclusiveness Audit subcategory provide an expository review of AI applications' inclusiveness by external experts, ensuring unbiased assessments. Impact Assessments, under Impact Assessment, offer a diminutive but significant glimpse into the real-world effects of AI on various groups. Algorithmic Transparency Reports from the Algorithmic Transparency subcategory elucidate AI decision-making processes, fostering trust and clarity. Accessibility measures ensure AI tools cater to users with disabilities, addressing a crucial aspect of inclusiveness. Lastly, Ethical Considerations, exemplified by User Control Over Data, ensure ethical compliance and user autonomy in data usage, upholding ethical standards in AI operations. The Unified Inclusiveness Score (UIS) formula is designed to quantify the inclusiveness of AI in project management and interactive systems. This formula synthesizes various measures into a single, comprehensive score, facilitating a structured and balanced evaluation.

To implement the UIS formula, each measure identified in the tables for Project Management AI and Interactive Systems with Live Data is assigned a specific weight $Weight_i$. These weights reflect the relative importance or relevance of each measure to the overall goal of inclusiveness. For example, User Satisfaction Surveys or Inclusiveness Audits might be assigned higher weights if deemed more critical in assessing inclusiveness. Each measure is then evaluated and given a score $Score_i$. This score can be derived from quantitative data, such as survey results or usage statistics, or qualitative assessments from expert reviews. The UIS is calculated by multiplying each measure's score by its weight and summing these products, *Unified Inclusiveness Score (UIS) = sum (Weight$_i$ * Score$_i$)*. This weighted scoring approach ensures that more crucial aspects of inclusiveness have a proportionately greater impact on the overall UIS. It offers a systematic way to assess and compare the inclusiveness of AI applications, ensuring that various dimensions of inclusiveness are considered. Key considerations in this approach include maintaining consistency in how measures are defined and applied, involving stakeholders in defining weights and interpreting scores, regularly updating the formula to reflect changes in technology and user needs, and validating the UIS with real-world data to confirm its effectiveness in measuring inclusiveness. The UIS formula, thus, serves as a valuable tool in guiding and evaluating the inclusiveness of AI in project management, ensuring that AI systems are not only efficient and effective but also equitable and inclusive.

4. Discussion and conclusion

Measuring inclusiveness necessitates a comprehensive approach, wherein a diverse set of criteria is established. These criteria, encompassing various dimensions of AI integration and functionality, are instrumental in assessing the impact of AI tools and systems on diverse groups within project management contexts. The methodology, while seemingly quixotic in its ambition to encapsulate the multifaceted nature of inclusiveness, adopts a categorical assertion of the importance of each criterion in contributing to the inclusive deployment of AI. The first dimension, focusing on technical aspects, includes criteria such as AI Decision Transparency, which underscores the need for clarity in AI decision-making processes. This clarity is not just perfunctory but vital in fostering trust among stakeholders, ensuring they comprehend the rationale behind AI-driven decisions. Algorithmic Bias Detection is another substantive criterion, where the emphasis is on identifying and correcting biases in algorithms, an ameliorative measure crucial for averting discriminatory outcomes. Data Diversity is highlighted for its role in training AI systems, advocating for the use of diverse datasets to enhance the system's inclusivity. Accessibility, ensuring AI tools are user-friendly for people with disabilities, and Language Support, catering to multilingual requirements, are also integral, reflecting the sublime goal of creating universally accessible AI tools. In the user interaction domain, criteria like User Interface Usability and Customization Options are pivotal. An intuitive interface, for example, enables seamless engagement with the AI tool, irrespective of the user's technical prowess. Customization Options allow users to tailor the AI tool to their individual needs, enhancing the tool's applicability. Feedback Mechanisms and Cultural Sensitivity are also crucial, ensuring that AI tools evolve in response to user needs and are sensitive to diverse cultural backgrounds. The methodology extends to evaluating specific project management aspects like Task Allocation Efficiency and Risk Assessment Accuracy. These criteria assess how effectively AI tools distribute tasks and evaluate project risks, respectively. Collaboration Enhancement and Timeline Management are also examined, evaluating AI's role in fostering team collaboration and managing project timelines. Ethical considerations form a significant part of the methodology. Criteria such as Privacy Protection and Security Measures are crucial in the digital age, where data confidentiality is paramount. Transparency in Data Usage and Ethical Standards Compliance ensure responsible and ethical AI usage. Performance metrics like Accuracy in Predictive Analytics and Response Time are evaluated to measure the AI tool's effectiveness. Organizational impact is assessed through criteria like Employee Satisfaction and Change Management, determining the broader effects of AI integration in the workplace. The third part of the methodology encompasses broader social, environmental, and stakeholder considerations. Social Impact, Environmental Sustainability, and Community Engagement are evaluated to ensure AI tools contribute positively to society and the environment. Innovation and Development focus on continuous improvement and adaptability of AI tools. Stakeholder involvement is emphasized, with criteria like Client Satisfaction and Investor Interest being assessed. This involvement ensures that the AI tool aligns with the needs and expectations of various stakeholders. The Unified Inclusiveness Score (UIS) formula serves as the capstone of this methodology. It quantifies the inclusiveness of AI in project management by synthesizing the measures into a comprehensive score. Each measure is weighted and scored, reflecting its importance in the overall inclusiveness. The UIS is calculated by multiplying each measure's score by its weight, offering a systematic way to assess and compare the inclusiveness of AI applications. This approach ensures that various dimensions of inclusiveness are considered, making the UIS a valuable tool in guiding and evaluating the inclusiveness of AI in project management. Regular updates and validations with real-world data are key to maintaining the formula's relevance and effectiveness. Thus, this methodology, through its perspicacious insight and assiduous execution, embarks on the laudable goal of ensuring that AI systems in project management are not only efficient and effective but also equitable and inclusive.

### 4.1. Limitations

The study, while comprehensive in its exploration of inclusiveness measures for PM models and datasets, encounters certain limitations that warrant acknowledgment. Primarily, the actual implementation of the suggested measures for enhancing inclusiveness within PM models and datasets, including assessing their performance and realistic applicability, represents a significant challenge. The measures proposed herein are contingent upon various factors, including the specific requirements and constraints of the business industry, which may not be universally applicable across all sectors. Hereupon, it is essential to recognize that these measures cover general aspects of the business industry, which are inherently limited in scope. The fluctuating and dynamic nature of other industries means that the applicability and effectiveness of these measures may vary considerably. Relative to the business industry, other sectors might require distinct approaches and modifications to these inclusiveness measures to align with their unique operational dynamics and challenges. Additionally, the focus of the study is akin to covering measures for basic PM ceremonies and task types, which, while fundamental, do not encompass the entire spectrum of PM activities. This focus, wherewith it provides a solid foundation, may not fully capture the complexities and variations inherent in more advanced or specialized PM tasks. The limitations of this study, therefore, lie in its scope, which is primarily tailored to standard business practices and basic PM functionalities. Hereafter, future research should aim to expand the scope of these measures, simultaneously exploring their applicability and effectiveness in a broader range of industries and for a more diverse set of PM tasks. This expansion is crucial for developing a more comprehensive understanding of how inclusiveness can be effectively integrated into PM across various sectors and for different types of tasks. By doing so, the field of PM can evolve to become more inclusive, efficient, and adaptable, meeting the diverse needs and challenges of the modern professional landscape.

### 4.2. Future Research

The research area of AI inclusiveness, brimming with potential, beckons further exploration, particularly in the realms of interactive feedback filtration from online sources and domain-specific applications. The future work in this field is poised to make significant strides in mitigating biases, thereby enhancing the effectiveness and inclusivity of AI tools in project management. The opportunity to investigate the interactive filtering of live feedback from online sources presents [Karthikeyan, Sekaran, Ranjith, Balajee, and others(2019)] an intriguing avenue. This approach can facilitate the avoidance of biases by dynamically adjusting the AI models based on real-time user input and diverse perspectives. It could lead to the development of AI systems that are more adaptable and sensitive to the evolving needs and nuances of different user groups. Moreover, the exploration of how insular communities within corporations can contribute to avoiding biases in project management, without sacrificing performance and quality, is an area ripe for investigation. Therein lies the potential to harness the collective wisdom [Banerjee, Gang, and He(2023)] and varied experiences of these communities to refine AI tools, making them more equitable and efficient. This approach underscores the importance of involving diverse stakeholder groups in the AI development process, ensuring that their insights and concerns are integral to the AI system's evolution. Conversely, the involvement of domain experts in industries like healthcare to experiment with AI in project management offers another promising opportunity [Nieto-Rodriguez and Vargas(2023), Stoumpos, Kitsios, and Talias(2023)]. For instance, the role of clinical project managers can be transformed through AI integration, enhancing decision-making processes and operational efficiency. In such settings, AI can provide invaluable assistance in managing complex projects, balancing various factors like regulatory compliance, patient safety, and research objectives. After all, the realm of AI in project management is not limited to traditional domains. There is, ipso facto, a plethora of opportunities awaiting exploration in areas like financial simulations co-piloting. These tools can provide real-time, on-the-fly analysis and predictions, aiding project managers in making more informed financial decisions. The potential of these AI applications is vast, with their hyperparameter tuning and eigenvalue calculations offering nuanced insights into financial scenarios. On the contrary, excluding these innovative AI applications from future research would be a missed opportunity to revolutionize project management across various industries. Hence, it is imperative to delve deeper into these prospects, exploring how AI can be tailored to specific domain needs while maintaining a commitment to inclusiveness and bias mitigation. Therefore, the future work in the realm of AI inclusiveness in project management should focus on harnessing the power of AI to cater to diverse needs, ensuring equitable access and representation. This entails not only the technological development of AI tools but also a nuanced understanding of the ethical, social, and practical implications of AI in diverse contexts. By doing so, AI can truly become a tool that enhances efficiency, fairness, and inclusivity in project management, thereby contributing to a more equitable and effective professional field.